\documentclass[conference]{IEEEtran}
\IEEEoverridecommandlockouts
\usepackage{cite}
\usepackage{amsmath,amssymb,amsfonts}
\usepackage{graphicx}
\usepackage[caption=false,font=footnotesize]{subfig}
\usepackage{algorithm}
\usepackage{braket}
\usepackage[noend]{algpseudocode}
\usepackage{textcomp}
\usepackage{xcolor}
\usepackage{geometry}
\def\BibTeX{{\rm B\kern-.05em{\sc i\kern-.025em b}\kern-.08em
    T\kern-.1667em\lower.7ex\hbox{E}\kern-.125emX}}
\begin{document}

\title{Quantum-Efficient Reinforcement Learning Solutions for Last-Mile On-Demand Delivery\\ 
{\large \textsuperscript{*}{\\
		\textbf{2025 IEEE Quantum Artificial Intelligence (QAI), Naples, Italy, November 2025.}\\
		{Funding source: Natural Sciences and Engineering Research Council \& Canada Research Chair}
}}
}

\author{\IEEEauthorblockN{Farzan Moosavi}
\IEEEauthorblockA{\textit{Laboratory of Innovations In Transportation} \\
\textit{Toronto Metropolitan University}\\
Toronto, Canada \\
0009-0008-8490-3851}
\and
\IEEEauthorblockN{Bilal Farooq}
\IEEEauthorblockA{\textit{Laboratory of Innovations In Transportation} \\
\textit{Toronto Metropolitan University}\\
Toronto, Canada\\
0000-0003-1980-5645}}

\maketitle

\begin{abstract}

Quantum computation has demonstrated a promising alternative to solving the NP-hard combinatorial problems. Specifically, when it comes to optimization, classical approaches become intractable to account for large-scale solutions. Specifically, we investigate quantum computing to solve the large-scale Capacitated Pickup and Delivery Problem with Time Windows (CPDPTW). In this regard, a Reinforcement Learning (RL) framework augmented with a Parametrized Quantum Circuit (PQC) is designed to minimize the travel time in a realistic last-mile on-demand delivery.  A novel problem-specific encoding quantum circuit with an entangling and variational layer is proposed. Moreover, Proximal Policy Optimization (PPO) and Quantum Singular Value Transformation (QSVT) are designed for comparison through numerical experiments, highlighting the superiority of the proposed method in terms of the scale of the solution and training complexity while incorporating the real-world constraints.

\end{abstract}


\section{Introduction}

Quantum Computation (QC) paradigms reveal the capability to bypass the classical computation barriers and obtain exponential speedup by quantum properties such as superposition and entanglement \cite{b1}, especially signifying great potential in solving transportation systems problems, such as Vehicle Routing Problem (VRP) \cite{b2}. 


Numerous works have offered the superiority of their quantum-based approach over current classical counterparts \cite{b3, b4, b5}, in terms of scalability, performance, and computational cost. However, when it comes to a large-scale graph at a city-level solution as well as incorporating realistic constraints, the quantum pipeline requires additional layers to account for such limitations. 
In this regard, recent studies have indicated the early promise in the following subjects to overcome such challenges. 
A significant implication of solution feasibility is how the problem is formulated, and through which mathematical notation, decision variables and constraints are encoded to qubits \cite{b6}.
To this end, Variational Quantum Algorithm (VQA) and a specialized case, Quantum Approximate Optimization Algorithm (QAOA), have been adopted widely due to their flexibility in circuit design \cite{b5, b6}. It involves matching the form of the circuit ansatz to the problem Hamiltonian. However, such a specific ansatz requires large circuit depths, making the circuit too expressive \cite{b7}, which often leads to barren plateaus, regions of the optimization landscape with vanishing gradients. 

In addition, the majority of such improvement is owing to the quantum algorithm and Hamiltonian simulation \cite{b7}, which represents the problem cost function. Linear Combination of Unitaries (LCU) \cite{b8}, and Quantum Singular Value Transformation (QSVT) \cite{b9}, in addition to QAOA trotterization, have been introduced. The former two have shown that the circuit complexity can be reduced from $O(N^2)$ in the case of QUBO-QAOA to $O(N)$ and $O(log(N))$ for LCU and QSVT, respectively \cite{b10}. The former presents a linear approximation of the Hamiltonian, which fails to capture the higher-order and non-linear pattern of the cost function. Whereas, the QSVT provide richer interactions between qubits by using polynomial transformations 
Nevertheless, they should be defined for the new optimization problem that can handle constrained VRP problems.


On the other hand, parametrized Quantum Circuit (PQC) design in combination with Reinforcement Learning (RL) demonstrated a promising approach \cite{b11, b12} due to an efficient encoding scheme and exploration via quantum neural network and the ability to manage a high-dimensional state-action space. For example, one qubit per node in the graph is introduced using Q-learning, where the Q-function is represented by a PQC to solve simplified VRP \cite{b13}. Moreover, the PQC can be designed such that it balances expressivity and trainability to outperform the classical methods. Nevertheless, the literature in this field of methodology needs more exploration to address real-world VRP with complex constraints.

\begin{figure*}[h]
\centerline{\includegraphics[width=0.7\textwidth]{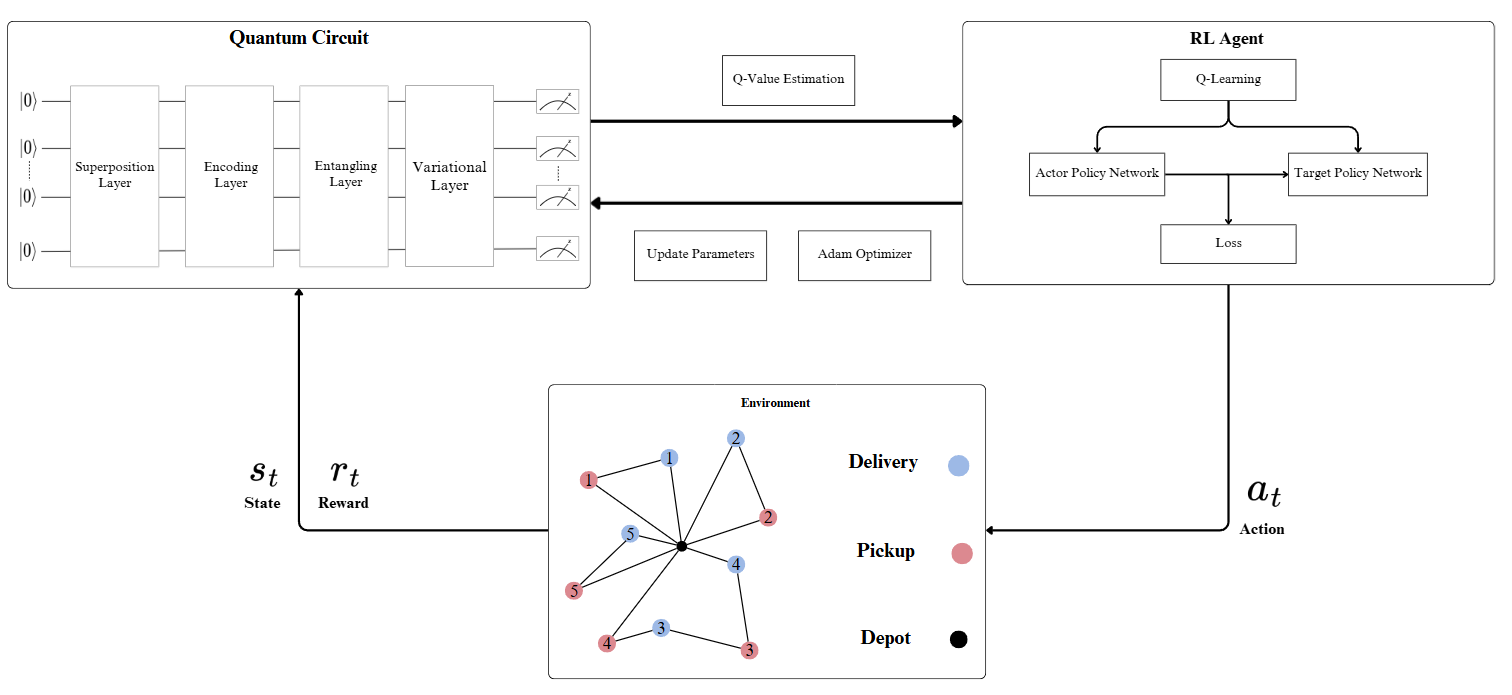}}
\caption{Overview of quantum reinforcement learning (QRL).}
\label{train}
\end{figure*}

Hence, this research proposes quantum RL optimization for on-demand last-mile delivery by autonomous vehicles to optimize the travel time.  This is modelled as one of the vehicle routing problems, the Capacitated Pickup and Delivery Problem with Time Windows (CPDPTW)  variants, constrained by the vehicle capacity, time windows, and node precedence constraints of pickup and drop-off of nodes. The solution framework is shown in Fig. \ref{train}.

In summary, our contribution is as follows:

\begin{itemize}
    \item Novel quantum-learning on-demand delivery optimization addressing real-world constraints;
    \item Problem-specific circuit design for quantum Hamiltonian approximation techniques;
    \item Scaling the solution and performance improvement with fewer trainable parameters. 
\end{itemize}

\section{Literature}

In this section, quantum RL-based literature in VRP is explored. RL can be combined with quantum algorithms in various ways. 
For example, a less memory-intensive approach than deep neural networks for dynamic task assignment was proposed using integration of PQC as function approximators for RL policies \cite{b14}. Rather than integrating the RL into the inner structure of quantum algorithms, it can be done in a reverse way, such as the research of \cite{b15}, which investigated finding the minimum number of SWAP gates to route a logical quantum circuit to a quantum computer formulated as a sequential decision-making process by employing reinforcement learning (RL) and Monte Carlo Tree Search (MCTS). In other words, in this work, the Markov decision variables are defined as circuit features like logical gates remaining, actions are possible SWAPs on physical edges, and the reward is the number of gates to be scheduled after SWAP. 

Another way to do this is to utilize the RL features in the quantum circuits, such as neural networks, to estimate the quantum parameterized circuit using different quantum feature maps to encode the input dataset. \cite{b16} proposed a hybrid quantum-classical RL method to address the split-delivery vehicle routing problem (SDVRP), and their key innovation lies in replacing classical multi-head attention mechanisms within a policy-based reinforcement learning framework. Even though the quantum advantage was not completely seen, they 
observed that selected non-linear observables, which were used to infer attention weights in a quantum circuit, would not
traditionally be considered in machine learning models, and it was not harmful to the policy performance. Therefore, they suggest new creative layers that could be
used in machine learning models. \cite{b17} introduced Quantum Neural Network (QNN) to solve constrained combinatorial optimization problems like Travelling Salesperson Problem (TSP). They proposed a learning-based, general-purpose, permutation-invariant quantum model that incorporates constraints directly into the circuit with high accuracy and low runtime. However, the formulation encoding was not suitable for scalability since it was of a complexity $O(N^2)$.
On the other hand, \cite{b18} presented a quantum Q-learning model that employs a PQC within a Deep Q-Learning Network (DQN) framework to solve the CVRP.
They have used a more expressive yet qubit-efficient PQC to embed dynamic state variables (like demand, load, and visitation) with significantly fewer trainable parameters and qubits of a complexity $O(N)$. They achieved a comparable result to classical DQN with many fewer trainable parameters. 
Additionally, \cite{b19} studied Equivariant Quantum Circuits (EQCs) within Quantum Graph Neural Networks (QGNNs) to solve the TSP, leveraging quantum reinforcement learning to train the circuit while capturing the symmetry of the problem. Their proposed model is shown to sometimes outperform the Christofides algorithm and is more promising due to their ability to handle problem-specific features like node-edge connectivity. 
However, the full potential for quantum advantage in these settings is still an open area of research.

In conclusion, learning-based methods present better performance in terms of the quality of the solution and complexity in comparison to classical ones, although scalability and generalization to larger and different problem scopes still need to be improved.

\section{Methodology}

This section initially explains the problem definition and environment setting, followed by how the optimization problem is solved through various approaches. 

\subsection{Problem Description}\label{AA}

The CPDPTW problem and constraints related to on-demand delivery are formulated as a Quadratic Unconstrained Binary Optimization (QUBO) problem.  
The QUBO notation is built upon the work of \cite{b5} for capacitated VRP, which accounts for time windows and precedence constraints. First, the Hamiltonian objective is shown in Equation \ref{eqobj}. 

\begin{equation}
H_{\mathrm{obj}}=\sum_{i, j, t, k} w_{i j} \cdot x_{i, t}^k \cdot x_{j, t+1}^k   \label{eqobj}
\end{equation}

Where $w_{ij}$ is the time travelled by vehicle $k$ between node $i$ and $j$, also $x^k_{i,t}$ is a binary decision variable, indicating whether the vehicle $k$ visits vertex $i$ at a time step $t$.
In this problem, electric vehicles (EVs) operate in a directed graph. The nodes are represented by network $G=(N, E)$, where $N=\left\{1, \ldots, n\right\}$ denotes the set of customers, and $E=$ $\left\{(i, j) \mid 0 \leq i, j \leq n, i \neq j \right\}$ denotes the set of edges connecting the nodes. The $i$th request demand and late time window are denoted by  $q_i$ and $l_i$, respectively, with $q_i>0$ for pickup nodes. The vehicles set is $k \in \left\{1, \ldots, K\right\}$, and each has a capacity $Q_k$. 
Subsequently, the constraints must be incorporated into problem formulation, as demonstrated through Equation \ref{eq1} through \ref{eq6}.

\begin{equation}
\sum_{k=1}^{K} \sum_{t=1}^T x_{i, t}^k=1 \quad \forall  i \in N 
    \label{eq1}
\end{equation}

\begin{equation}
\sum_{t=1}^T \sum_{i=1}^n x_{i, t}^k \cdot q_i \leq Q_k \quad \forall k \in K
    \label{eq3}
\end{equation}

\begin{equation}
\sum_{t=1}^{j} \sum_{i,j = 1}^{n} w_{ij} \cdot x_{i, t}^k \cdot x_{j, t+1}^k \leq l_j \quad \forall k \in K
    \label{eq5}
\end{equation}

\begin{align}
 \tau(i)<\tau(j)   \rightarrow 
 \sum_{t=1}^{i} x_{i, t}^k  =
 \sum_{t=i+1}^{T} x_{j, t}^k 
 \quad \forall(i, j) \in P,   \forall k \in K
    \label{eq6}
\end{align}

Equation \ref{eq1} satisfies visiting each customer exactly
once. Equation \ref{eq3} represents the capacity constraint of each vehicle. We define Equation \ref{eq5} based on the time variable to account for the time window. Finally, Equation \ref{eq6} accounts for the precedence constraint. $\tau(i)$ returns the ordering of node $i$ in the considered sequence. $P$ is the set of all precedence constraints such that if $i$ has to precede $j$, this is represented by the pair $(i, j)$ and then their corresponding node must be visited chronologically. 

\subsection{Quantum Reinforcement Learning}\label{AA}

\subsubsection{DQN}

The RL method we have adopted is Deep Q-Learning Network (DQN), where a customized PQC is replaced with a classical neural network. DQN is defined in the scope of Markov Decision Process (MDP). A generic MDP consists of four components: state space $S$, action space $A$, reward function $r(s, a)$, and state transition probability $p\left(s^{\prime} \mid s, a\right)$. 

\begin{itemize}
    \item \textit{States:} Composed of the graph vertex state and the vehicle state, denoted as $s_t ={\left\{x_t, v_t\right\}} $ at step t respectively, where $x_t = (x,q_t,l_t)$ and $v_t = (u_k,t_k)$. $x$ is the two-dimensional coordinate for location, $q_t$ is the demand, and $l_t$ is the late-time window for each node. $u_k$ and $t_k$ are the load and time starting from the depot for vehicle $k$. 

    \item \textit{Actions:} $a_t$ determines the vehicle's node selection at step t. The sequence of actions generated from the initial to the final step should be a combination of nodes starting and ending with the depots.

    \item \textit{Reward:} PDPTW aims to minimize fleet delay and travel time. Therefore, the reward function is given in Equation \ref{eqre}.

\begin{equation}
 \; R=\alpha_1\sum_{k \in K} \sum_{(i, j) \in N} t^k_{i j}  +\alpha_2\sum_{i \in P \cup D}  max\left\{T_{k}-l_i, 0\right\}
    \label{eqre}
\end{equation}

\end{itemize}

Where $\alpha_1$ and $\alpha_2$ are the monetary coefficient factors accounting for EV transportation time cost and the time delay cost, respectively. Besides, $t^k_{ij}$ denotes the time taken for vehicle $k$ to travel from node $i$ to $j$.

DQN is an approximation method to estimate the Q-value function for the given possible states and actions. It represents the expected cumulative reward for each possible action $A$ at a given state $S$. At time step $t$ of the total time $T$, the Q-value is defined as Equation \ref{q}.

\begin{equation}
    Q^\pi\left(a_t, s_t\right)=E\left[r_t+\sum_{t \prime=t+1}^T \gamma^{t \prime-t} r_{t^{\prime}}\right]
    \label{q}
\end{equation}

where $r$ is the feedback reward and $\gamma$ is the discount factor, accounting for future reward in each time step. Furthermore, $\pi$ refers to the policy, a strategy for making the decision according to Q-values. In DQN, the data for training is collected by interacting with the environment, observing the environment state $s$, choosing an action $a$, which turns the state into $s^{\prime}$ and obtaining a reward $r$. The experience ( $s, a, r, s^{\prime}$ ) is stored in a fixed-length replay buffer, periodically sampled to train the Q-network \cite{b18}. Initially, the interaction to fill the experience buffer memory is a random node assignment to complete the delivery and the initial Q-network. Afterwards, through iteration, the policy employs the epsilon-greedy to explore optimality for each batch, which is sampled from the memory. The resulting Q-values will then be updated into the target Q-network using the minimizing the loss function, and the Q-network parameters will be updated accordingly by the loss formula in Equation \ref{loss}.

\begin{equation}
    \mathcal{L}(\theta)=E_{\left(s, a, r, s^{\prime}\right) }\left[\left(r+\gamma \max _{a^{\prime}} Q_{\theta^{\prime}}\left(a^{\prime}, s^{\prime}\right)-Q_\theta(a, s)\right)^2\right]
    \label{loss}
\end{equation}

It is noted in the sufficient episode that for a certain amount of data, the target network will gradually reach optimal parameters, $\theta^{\prime}$, which is then periodically copied from those of the Q-network. The overall quantum DQN is shown in Algorithm \ref{alg1}. 

\subsubsection{PQC}

\begin{figure*}[h]
\centerline{\includegraphics[width=0.7\textwidth]{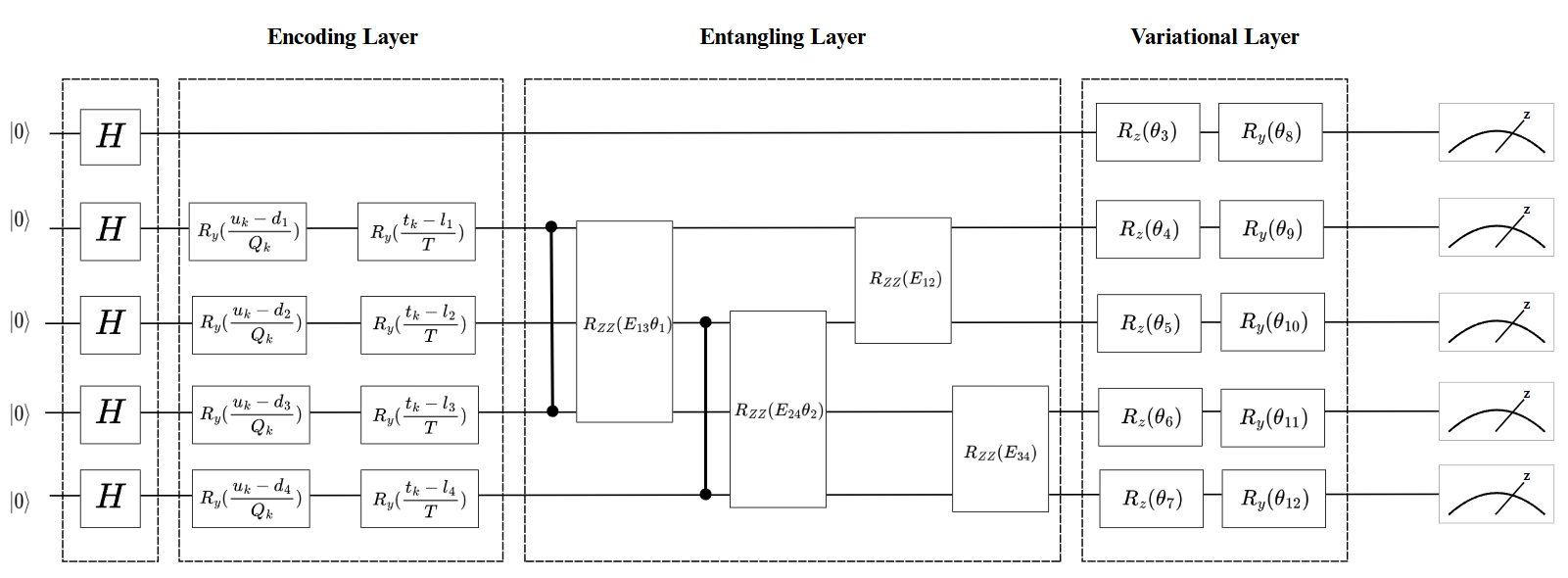}}
\caption{Parametrized circuit design stages for a special case of N=5. Starting with the superposition layer to excite the parallel computation for each qubit and angle encoding of instance features, namely, vehicle and node states. The relative load and time of the vehicle are encoded for each node separately with a Y-rotation gate. Subsequently, to take the pickup and delivery precedence, the two entangling operations are adopted, controlled-CNOT gate followed by a learnable spatially-sensitive ZZIsing feature map. Finally, the variational layer enhances the expressivity of the circuit while being moderate in complexity.}
\label{1}
\end{figure*}

The quantum circuit, as a shortcut, is designed to perform the q-value estimation for the constrained on-demand last-mile delivery problem. The PQC is structured into several distinct layers, each playing a specific role in encoding classical data, capturing problem features, relations, and extracting trainable quantum features. Note that in this architecture, the number of qubits is equal to the number of nodes; therefore, the scale complexity is $O(N)$. Extracting non-linear and hidden interaction of the nodes and trade-off between circuit expressivity and trainability, this operation will be done in $p$ layers. Initially, all qubits put in a superposition state using a Hadaramrd gate, transform into $|+\rangle^{\otimes(N+1)}$. Next, to convert environment features into the quantum state, dynamic encoding of real-time state data, load and time of the vehicle with respect to the vehicle capacity and late time window, in that order. First, a Y-rotation is used to both encode the normalized load and time, such as $\frac{(u_k-q_i)}{Q_k}$ and similarly, $\frac{(t_k-l_i)}{T}$ for each vehicle and demand for node $i$. $T$ is the total time frame of the delivery.
Afterwards, to highlight the pickup and delivery coupling strength and correlation, a controlled-Z-rotational entangling layer for each pickup to its corresponding delivery node, followed by a distance-based IsingZZ entangling layer between every pickup node and another between every delivery node. Additionally, to ensure the flexibility of the PQC to estimate the Q-value function, two parametrized rotational encodings of Z-rotation and Y-rotation, respectively, are applied in the variational layer. Finally, to obtain classical features from a quantum state, the expectation values of the Pauli-Z measurement are calculated for each qubit to feed into a classical linear layer. To clarify, each layer of the PQC has a unique set of trainable parameters, and weights to be trained are defined in the angle encodings in the variational layer, and in the CRZ entangling layer to account for the pickup and delivery connection, for each qubit. Therefore, for one layer of PQC and a graph of size N, there would be $2N+\frac{N}{2}$ trainable parameters for pickup and delivery pairs, and another 2 for the depot, encouraging the depot revisit for vehicles. Besides, a data reuploading scheme is augmented to the circuit, which will repeat the variational layers through another $p$ sequential layers, though with different learning parameters. 
A detailed PQC design for a simple case of two pairs of pickup and delivery is demonstrated in Fig. \ref{1} as an example.

\begin{algorithm}[h]
\tiny
\caption{Quantum Deep Q-Learning for CPDPTW}
\label{alg1}
\begin{algorithmic}[1]
\State \textbf{Input}: Number of requests $n$, vehicle capacity $C$, time window frame $T$
\State \textbf{Initialization}:
\begin{itemize}
    \item Generate city coordinates for depot, pickups, and deliveries
    \item Assign demands: pickups $d_i \in [1,3]$, deliveries $-d_i$
    \item Generate pickup and delivery time windows $[t_i^p, t_i^d]$
    \item Initialize environment state: vehicle at depot, zero load and time
\end{itemize}

\State \textbf{Initialize Quantum Q-Networks}: $\pi_\theta$ (policy), $\pi_{\theta'}$ (target), memory buffer $\mathcal{M}$

\For{each episode}
    \State Reset environment and get initial state $s_0$
    \While{not all requests visited}
        \State Select action $a_t = \arg\max Q_\theta(s_t, a)$ over feasible actions
        \State Execute $a_t$, observe $r_t$, $s_{t+1}$, receive feedback
        \State Store transition $(s_t, a_t, r_t, s_{t+1})$ in $\mathcal{M}$
        \State Sample batch from $\mathcal{M}$ and compute temporal difference targets
        \State Update $\pi_\theta$ using quantum circuit gradients
        \State Soft update target network $\pi_{\theta'} \leftarrow \tau \pi_\theta + (1-\tau)\pi_{\theta'}$
        \State $s_t \leftarrow s_{t+1}$
    \EndWhile
    \State Store total reward
\EndFor
\State \textbf{Output}: Trained policy $\pi_\theta$ for routing decisions
\end{algorithmic}
\end{algorithm}

\subsection{Hamiltonian QSVT Simulation}
 
To simulate Hamiltonian, the QSVT circuit is combined with the classical optimization loop and is parametrized with efficient gates and certain layers. For example, parameters of $\gamma$ and $\beta$, are optimized in a closed-loop using gradient descent, as shown in Equation \ref{loop}. 

\begin{equation}
\begin{aligned}
& \left({\gamma}^*, {\beta}^*\right)=\underset{{\gamma}, {\beta}}{\operatorname{argmin}} \; E({\gamma}, {\beta}), \\
& E({\gamma}, {\beta})=\langle{\gamma}, {\beta}| H_{{C}}|{\gamma}, {\beta}\rangle .
\end{aligned}
    \label{loop}
\end{equation}

This will be updated as more parameters are involved in the Hamiltonian and will iteratively run over the circuit to obtain the optimal parameters. 

\begin{figure*}[h]
     \centering
      \subfloat[Block-encoding circuit for three-node graphs.]{\includegraphics[width=0.4\textwidth]{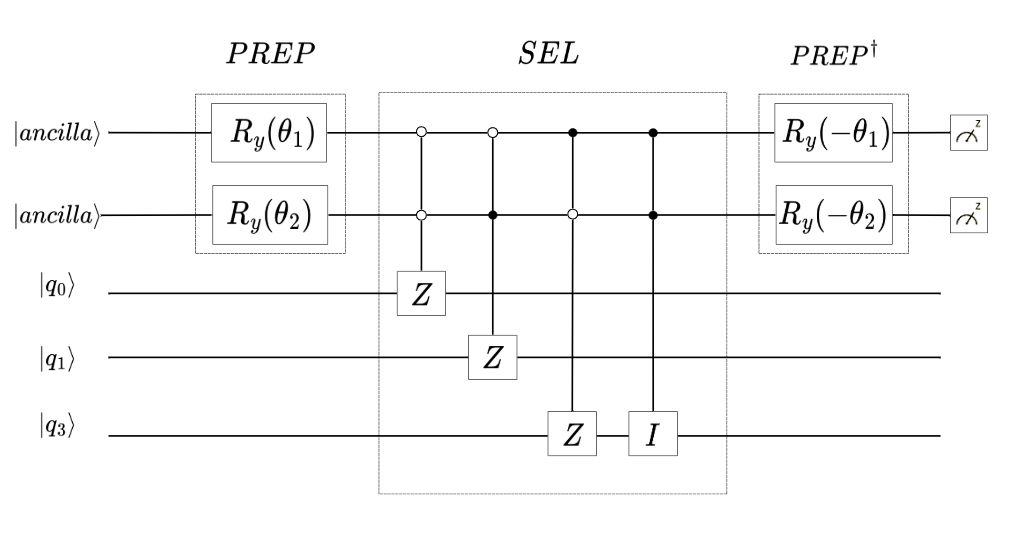}\label{cir1}}
     \hfill
      \subfloat[QSVT circuit for three-node graphs.]{\includegraphics[width=0.5\textwidth]{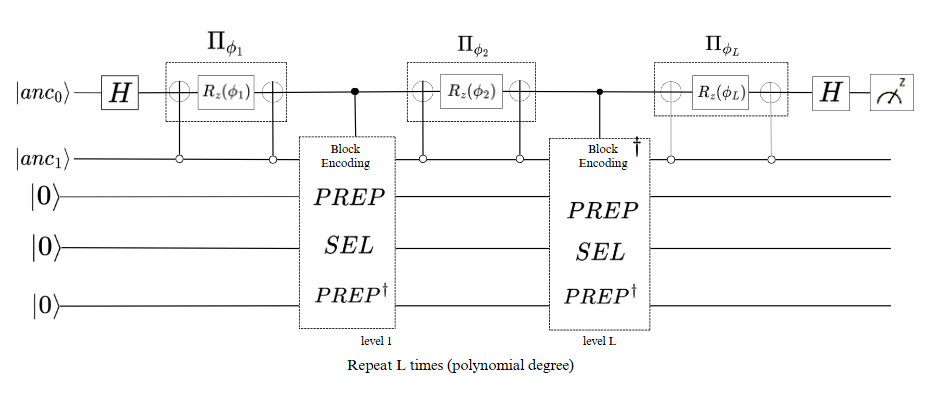}\label{cir2}}   
      \caption{Quantum circuit design for various Hamiltonian simulations.}
      \label{qs}
\end{figure*}

The QSVT is used to apply a polynomial transformation on the singular values of a block-encoded matrix (shown in Fig. \ref{cir2}). Similar to LCU, by alternating a block-encoding unitary and projector-controlled phase gates to the polynomial transformation for the Hamiltonian. 









 However, unlike LCU, QSVT incorporates generalizing the transformation to higher dimensions for better approximation. Given a unitary $U$, a list of phase angles and two projector-controlled-not operands, the QSVT algorithm is defined in terms of Hamiltonian singular value decomposition, which is shown in Equations \ref{qsvt} and \ref{qsvt1}, for even and odd polynomial degree  
$d$, respectively.

\begin{equation}
\left[\prod_{k=1}^{d / 2} \Pi_{\phi_{2 k-1}} U(A)^{\dagger} \tilde{\Pi}_{\phi_{2 k}} U(A)\right] \Pi_{\phi_{d+1}}
    \label{qsvt}
\end{equation}

\begin{equation}
\tilde{\Pi}_{\phi_1}\left[\prod_{k=1}^{(d-1) / 2} \Pi_{\phi_{2 k}} U(A)^{\dagger} \tilde{\Pi}_{\phi_{2 k+1}} U(A)\right] \Pi_{\phi_{d+1}}
    \label{qsvt1}
\end{equation}

$\Pi_{\phi_i}$ can be constructed using an auxiliary qubit, two CNOT gates, and a $Z$ phase gate. The circuit architecture for a specific case for QSVT is depicted in Fig. \ref{qs}.

\section{Experimental Simulation}

This section discusses the experiments and the results of the CPDPTW problem.
Specifically, the numerical simulations for quantum RL and classical optimization and the quantum circuits designed for the Hamiltonian evolution via QSVT are demonstrated.

\subsection{Initial Setting}\label{SCM}

The problem is initialized with node locations, which are drawn uniformly from the square area of $[0,1]km$, and the random delivery time window from $[20, 40]min$ based on the node. The maximum speed of the vehicles is set at $20 m/s$. The demand volume of a pickup node $d_i$ is uniformly sampled from $(1,3)$, and the capacity limit of each vehicle is 5. The monetary coefficient is chosen as $\alpha_1 = 0.6$ and $\alpha_2 = 0.05$. 

\subsection{Result Analysis}

To ensure the feasibility of the solution to avoid sparse reward training and avoid hard constraints, the time window violation is relaxed, meaning that the vehicles are allowed to visit outside the window, although the infeasible actions, such as node revisit, overload capacity, and pickup and delivery precedence, are masked. Besides, to enhance the learning performance due to the high-dimensional dataset and constraints, a prioritized sampling or importance sampling is used to store higher temporal differences in loss transitions with higher priority. Besides, a double DQN (DDQN) is used to reduce overestimation bias for better Q-value prediction as well as the epsilon-greedy decay over the episodes. During the training process, it was observed that, potentially due to the highly non-linear input feature, the DDQN failed to find the optimal solution and the convergence criteria were not met even with a 256-dimensional neural network embedding. In this regard, the pickup and delivery hard constraints were also relaxed instead of returning a constant penalty, as well as a PPO is proposed to use the policy gradient method, which can handle constraints more efficiently due to the log probability gradients and distribution over actions.

\begin{figure}[h]
\centerline{\includegraphics[width=0.5\textwidth]{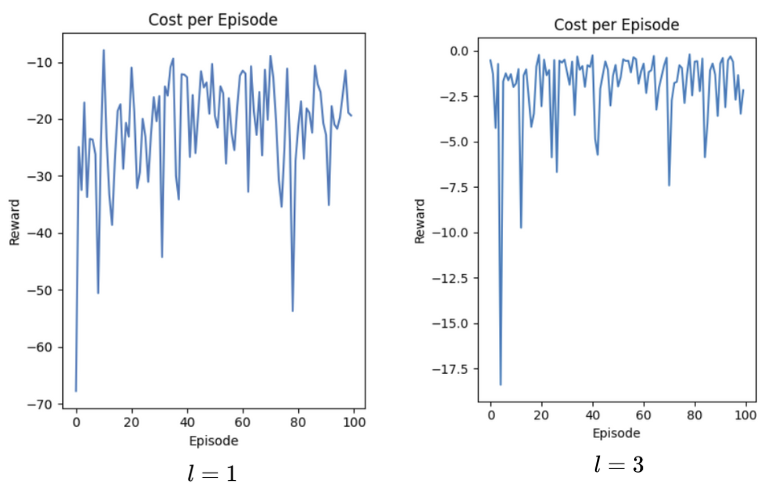}}
\caption{Training and loss learning curve for customized PQC-enhanced RL for two different PQC repetitive layers.}
\label{qrl}
\end{figure}

On the other hand, QSVT is proposed as another quantum baseline, which uses a mixer and a cost layer optimized with COBYLA. Fig. \ref{qrl} shows the training curve for the proposed QRL method for 100 episodes for the problem size of 10. The learning rate for the experiment is $\alpha=0.0005$, and the data size is considered 10000 instances. The learning will lead to lower average cost and faster convergence as the number of layers increases. For $p=3$, the curve is more stable and can find an optimal policy after 20 epochs, whereas for $p=1$, it requires at least 80 epochs to converge to a stable reward. This is due to the data reuploading feature of repetitive layers of PQC, leading to efficient and expressive circuit representation.

\begin{figure}[h]
\centerline{\includegraphics[width=0.5\textwidth]{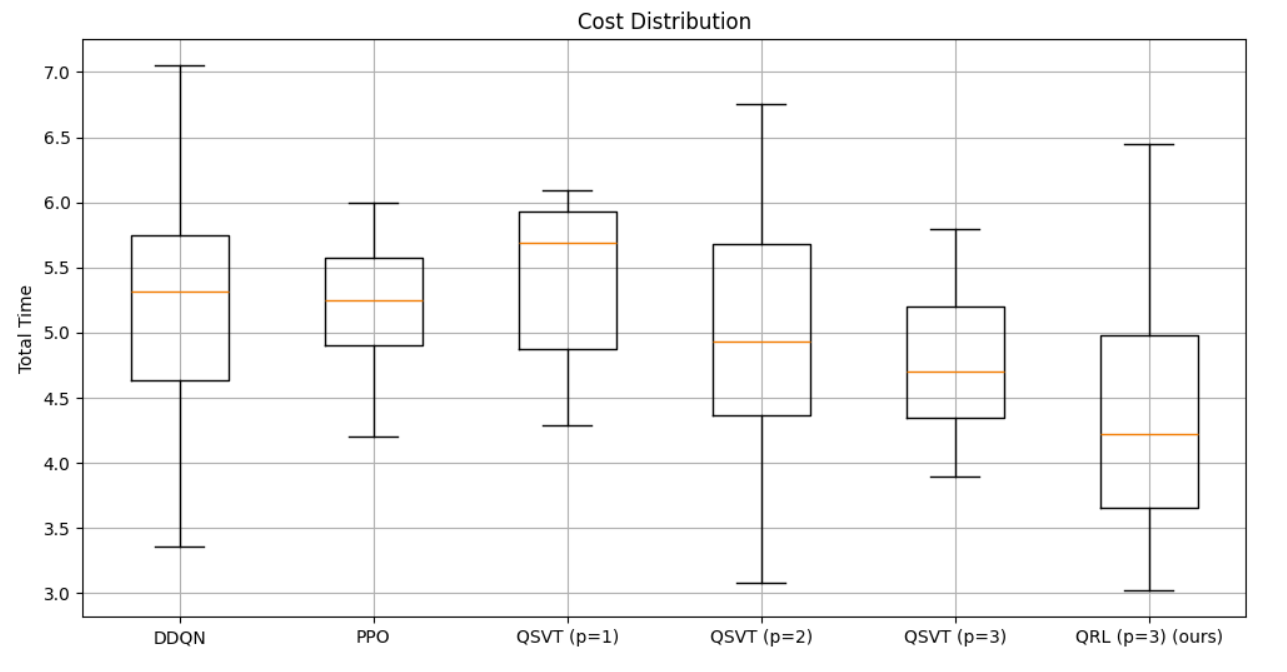}}
\caption{Total time distribution for various reinforcement learning methods applied to the CPDPTW problem.}
\label{gen}
\end{figure}

In addition, Fig. \ref{gen} illustrates the test instances' results using the trained model for the baselines in a box plot distribution of the problem cost. As can be seen, DDQN exhibits the highest variance due to its difficulty handling feasibility constraints in discrete action spaces, and PPO improves stability and reduces average cost due to entropy-based exploration. QSVT shows improved performance as circuit depth (p) increases owing to the polynomial transformation functions. Notably, our approach outperforms all others, achieving the lowest median cost by integrating entanglement-aware layers and combining quantum-inspired models with variational layers.

\section{Conclusion}

In this research, a novel parametrized quantum circuit scheme is introduced to tackle Capacitated Pickup and Delivery with Time Windows (CPDPTW). All quantum experiments were conducted on Pennylane quantum emulator running on a classical computer. Due to resource limitations, the results yield for small case problems; however, our proposed method could achieve superior performance to optimize the real-world constrained on-demand delivery for given extremely lower trainable parameters. 

In future direction, we focus on developing a more complicated reinforcement learning for graph instances, such as quantum decoding in the transformer architecture. Furthermore, circuits based on  Grover search are worth investigating for the constraint-specific problems in combination with the multi-head attention architecture to encode the attention scores and find the optimal value. 




\end{document}